# Photoswitchable exceptional points derived from bound states in the continuum

Lei Wang[1,†], Hang Liu[1,†], Junwei Liu[1], Aoxuan Liu[1], Jialiang Huang[1], Qiannan Li[1], Hui Dai[1], Caihong Zhang[1,2,*], Jingbo Wu[1,2], Kebin Fan[1,2,*], Huabing Wang[1,2], Biaobing Jin[1,2], Jian Chen[1,2], and Peiheng Wu[1,2]

[1]*Research Institute of Superconductor Electronics (RISE), Key Laboratory of Optoelectronic Devices and Systems with Extreme Performances of MOE, School of Electronic Science and Engineering, Nanjing University, Nanjing 210023, China*

[2]*Purple Mountain Laboratories, Nanjing 211111, China*

[†]*These authors contributed equally to this work*

*Correspondence should be addressed to: chzhang@nju.edu.cn and kebin.fan@nju.edu.cn.*

## Abstract

Bound states in the continuum (BICs) and exceptional points (EPs), as two distinct physical singularities represented by complex frequencies in non-Hermitian systems, have garnered significant attention and clear definitions in their respective fields in recent years. They share overlapping applications in areas such as high-sensitivity sensing and laser emission. However, the transition between the two, inspired by these intersections, remains largely unexplored. In this work, we reveal the transition process in a non-Hermitian two-mode system, evolving from one bound singularity to a two-dimensional exceptional ring, where the EP is the coalescent state of the quasi-Friedrich-Wintgen (FW)-BIC. This phenomenon is experimentally validated through pored dielectric metasurfaces in terahertz band. Furthermore, external pumping induced photocarriers as the dissipative perturbation, facilitates the breaking of degeneracy in the complex eigenfrequency and enables dynamic EP switching. Finally, we experimentally demonstrate a switchable terahertz beam deflection driven by the phase singularities of the EP. These findings are instrumental in advancing the development of compact devices for sensing and wavefront control within non-Hermitian systems.

# Introduction

The physical world often involves energy exchanges with the environment, often described by the framework of non-Hermitian physics. Consequently, the eigenmodes or quasi-normal modes of non-Hermitian systems are often represented by complex frequencies[1-3], where the imaginary parts indicate gain or loss. Among many interesting phenomena in non-Hermitian systems, bound states in the continuum (BICs) and exceptional points (EPs) are intriguing for their unique properties and potential applications. BICs refer to localized states embedded within the radiation continuum above the light cone, which has a zero imaginary component in its eigenvalue, theoretically resulting in an infinite quality (Q-) factor[4]. In contrast, an exceptional point (EP), where at least two modes in the eigenvalue spectrum collapse dimensionally with identical resonance positions and lifetimes, represents a critical phenomenon often associated with the parity-time (PT) symmetry. This EP serves as the phase transition point that delineates the boundary between PT symmetry and its breaking[5,6]. BICs and EPs have garnered significant attention in their respective fields and exhibit similar applications in sensing[7-12], laser emission[13-15], and unidirectional or non-reciprocal transmission[16,17].

Originated from quantum systems, studies on BICs and EPs have recently been actively explored in classical systems, such as photonics and optics[5,18-20]. This progress has also been facilitated by the development of artificial microstructures, including metamaterials and metasurfaces, which can be precisely engineered to manipulate electromagnetic waves[21-23]. Among these, dielectric metasurfaces stand out due to their low intrinsic material losses, making them an ideal platform for achieving a wide range of exotic scattering phenomena. For example, strong vibrational coupling has been achieved using quasi-BIC freestanding Si film metasurfaces[24], while the shift from Dirac points to exceptional rings in the shift from Hermitian to non-Hermitian systems has also been reported[25]. However, these studies focus separately on either BICs or EPs, without examining the interplay between the two. In the context of BICs, there exists one type referred to as Friedrich-Wintgen BIC (FW-BIC), where coupling resonance arises from the destructive interference of at least two modes[4,26-28]. Given the involvement of complex eigenfrequencies and multi-mode coupling—similar to those observed in EPs—this opens up the possibility of exploring the connection between these two phenomena. Although only a handful of studies have explored this topic, the focus has primarily been on the theoretical coexistence of BICs and EPs within a single system[29-31]. The merging of multiple BICs into a single EP, inheriting the hybrid characteristics of the two singularities, has also been numerically demonstrated[32]. Consequently, there is a pressing need for further research into the transition between these topological singularities, from BICs to EPs, as well as the active tunability of these phenomena.

In this work, we not only reveal the transition between two singular concepts in a non-Hermitian system, demonstrating that the FW-BIC can coalesce into an EP with a small imaginary part and a high quality (Q-) factor, but also experimentally observe this transition process by changing the incident wave vector angle and thus achieving in-plane wave vector matching using classical hollow dielectric metasurfaces. Moreover, we show that the degeneracy breaking of the system's EP can be achieved through carrier injection from external optical excitation, enabling a dynamic switching of the EP. Leveraging the phase singularities of the EP, we experimentally demonstrated the dynamic beam steering of terahertz beams through optical control. Our work presents a novel strategy for diverse phenomena in non-Hermitian singular systems, offering the potential to combine BIC and EP for applications in high-sensitivity sensing and wavefront manipulation of electromagnetic waves. Additionally, it could enable valuable insights for dynamic topological phase imaging and high-capacity vortex signal transmission.

## Results

The temporal coupled mode theory (TCMT) is a valuable tool for describing the underlying physics of such open systems[33], including near-field and far-field coupling between modes. First, we will theoretically show that BICs and EPs can simultaneously exist when at least two modes are present in an open system. For a simple two-mode model, the effective Hamiltonian is:

$$H_{\mathrm{eff}} = \begin{pmatrix} \omega_1 & \kappa \\ \kappa & \omega_2 \end{pmatrix} - i \begin{pmatrix} \gamma_1 & \gamma_{12} \\ \gamma_{21} & \gamma_2 \end{pmatrix}. \qquad (1)$$

The first term on the righthand is a lossless Hamiltonian with resonant frequencies $\omega_{1,2}$ of the modes and near-field coupling between the two modes $\kappa$, while the other term represents the non-Hermitian component that includes losses $\gamma_{1,2}$ for each mode and the far-field coupling term $\gamma_{12}$. If the system satisfies symmetry, $\gamma_{12}=\gamma_{21}=\sqrt{\gamma_1\gamma_2}$. By solving the eigenvalues of the matrix $H$, we can obtain the complex eigenfrequencies:

$$\omega_{\text{eig}1,2} = \frac{(\omega_1+\omega_2)-i(\gamma_1+\gamma_2)\pm\sqrt{[(\omega_1-\omega_2)-i(\gamma_1-\gamma_2)]^2+4(\kappa-i\sqrt{\gamma_1\gamma_2})^2}}{2}. \qquad (2)$$

Under the FW-BIC condition $\kappa(\gamma_1-\gamma_2)=\sqrt{\gamma_1\gamma_2}(\omega_1-\omega_2)$, one of the eigenfrequencies $\omega_{\text{eig}1}$ will be real, and the other one $\omega_{\text{eig}2}$ remains complex with a significantly large imaginary part. The detailed derivation is in Section 1 of the supplementary material. The positions of these eigenvalues and the gap between them depend on the sign and magnitude of the near-field coupling coefficient $\kappa$. Specifically, when $\kappa = 0$, the energy bands of the BIC and the lossy mode coalesce, i.e., Re($\omega_{\text{eig}1}$) = Re($\omega_{\text{eig}2}$) and Im($\omega_{\text{eig}1}$) + Im($\omega_{\text{eig}2}$) = Im($\omega_{\text{eig}2}$) >0. Further, the system can meet the EP conditions by tuning the losses of the eigenmodes, which is analogous to the experimental observation where the leakage of the BIC results in a quasi-BIC with a finite lifetime.

In momentum space, the Q-factor of BIC is typically inversely proportional to the square of the transverse momentum ($k$), meaning the loss is proportional to $k^2$, i.e., $\text{Im}(\omega_{\text{eig}1}) \propto k^2$. This indicates that by altering $k$, the mode loss can be tuned. And when the BIC mode is at $k = 0$, i.e., at the center of the first Brillouin zone (the Γ point of FBZ), increasing the $k$ leads to a symmetry mismatch, causing the BIC to transition into an observable quasi-BIC with a high Q-factor. At the same time, assuming that the total loss caused by the interference of the two modes forming the FW-BIC remains constant, the losses of the two modes can coincide at a specific transverse momentum, and their imaginary parts of the characteristic frequencies become equal, i.e., Im($\omega_{\text{eig}1}$) = Im($\omega_{\text{eig}2}$). As a result, the two modes coalesce into an EP with common features.

To validate our theoretical descriptions, we designed a silicon-based terahertz metasurface enabling the transition process from BIC to EP as shown in Fig 1a. The structural unit is composed of a low-loss perforated free-standing silicon slab with a radius of air holes r = 70.4 μm, arranged in a square lattice with a period of a = 230 μm. The silicon slab has a thickness of h = 120 μm. Because of the highly resistive silicon, extremely low material (non-radiative) loss rates can be achieved. Additionally, the advantage of operating at the terahertz region lies in the ability to modify the carrier density in silicon through optical doping, a process as we will demonstrate later, which facilitates the switching of the EP state. The coupling of modes is influenced by the structural parameters, and to illustrate this, we have selected the radius as a key parameter, with detailed results presented in Supplementary Material Fig. S1. Fig. 1b and c show the real and imaginary parts of the transverse electric (TE) like eigenmodes in $k$ space for the designed metasurface structure. The three colors represent three modes, with the white indicating an orthogonal mode that is disregarded. Our focus is on the red and blue dispersion curves, denoted $TE_1$ and $TE_2$, which correspond to the two modes forming the FW-BIC. At the $\Gamma$ point, the real parts of the two eigenvalues are degenerate. While one imaginary part close to zero indicates that it is a non-radiating state, i.e., BIC, the other has a larger loss, *i.e.* a lossy mode. As $k$ gradually moves away from the $\Gamma$ point to a critical value, the real and imaginary parts of the eigenmodes in each $k$-direction simultaneously degenerate, resulting in the emergence of an EP ring. This transition facilitates the evolution of the non-Hermitian system from a single BIC to a two-dimensional exceptional ring. The detailed dispersion curves of the two modes in the two-dimensional $k$ direction are shown in Fig. 2b. Two EPs in the $k$ directions from $\Gamma$ to X and M gradually divide the dispersion into two regions. Within the EP ring ($-0.023 < k < 0.023$), the real parts of the eigenfrequencies Re($\omega_{eig}$) are non-dispersive and degenerate, while outside the ring ($-0.023 > k$ and $k > 0.023$), they are dispersive and non-degenerate. The situations for the imaginary parts Im($\omega_{eig}$) are exactly opposite. From the PT symmetry perspective, $k$ inside the EPs indicates that the system is in a symmetry-broken phase, as opposed to those outside indicating a symmetric phase. The entire process is consistent with our previous TCMT analysis as shown in Fig S2 in the Supplementary Material. The band structure near the EP exhibits square root branching behavior, attributed to the consideration of non-Hermitian conditions such as open boundaries. The change in normal magnetic field distribution of the two modes from the $\Gamma$ point (BIC) to the off-$\Gamma$ bifurcation point (EP) reflects the gradual coalescence of the eigenstates. In addition, we also calculated the Q-factors of both modes in the FBZ and far-field polarization of BIC mode as shown in Fig. 2c and d. The Q-factor is defined by Q = Re($\omega_{eig}$)/2Im($\omega_{eig}$). It is noteworthy that the Q-factor of the $TE_1$ mode reaches its maximum value at the center of the FBZ, ideally approaching infinity, while the Q-factor of the $TE_2$ mode exhibits a minimum value. Interestingly, both modes show abrupt changes at the boundary of the EP ring. A winding behavior of far-field polarization can also be observed at the center of the FBZ, forming a polarization vortex. By calculating the number of windings of the polarization vector in the counterclockwise direction, the topological charge can be easily determined. The definition of the topological charge is as follows[34,35]:

$$q = \frac{1}{2\pi} \oint_c d\mathbf{k} \nabla_k \varphi(\mathbf{k}), \tag{3}$$

where $c$ represents a closed path in the $\mathbf{k}$-space, oriented in an anti-clockwise direction around the BICs, and $\varphi(\mathbf{k})$ denotes the angle between the polarization major axis and the $x$-axis. In our case, the BICs are all topologically protected by the same topological charge of +1, which is an important property of the topological nature of the BICs.

To experimentally demonstrate the theoretical and simulation findings, we patterned the structure on a thin (120 μm) and high-resistivity (1,000 Ω·cm) silicon wafer using standard photolithography and deep reactive ion etching processes. The left side of Fig. 2a presents the optical microscope image of the large-area sample, while the right side illustrates the scanning electron microscope (SEM) cross-sectional image of the sample, providing a detailed view of our fabricated sample. Since the designed FW-BIC is located at the Γ point, it can also be considered as a symmetry-protected BIC. The wave vector of the incident wave exhibits a sinusoidal relationship with the $\mathbf{k}$, and the transition from BIC to quasi-BIC can be observed by varying the incident angle for mode matching. A Terahertz time-domain spectroscopic system (THz-TDS) with a spectral resolution of about 1 GHz, is used for the characterization of the fabricated large-area metasurfaces. A stepper motor was installed on the transmitting and receiving antennas to adjust the incident angle of the THz waves with an angular resolution of 0.1°, ranging from 0 to 6.7°. This setup allows for the characterization of angle-resolved transmission spectra along the Γ-X direction. Detailed sample preparation and testing procedures can be found in the Methods. Figure 3a illustrates the measured transmission spectra as a function of the incident angle, alongside the extracted transmission profiles at marked incident angles of 0°, 1.5°, and 5°, corresponding to BIC, EP, and more lossy quasi-BIC, respectively. These results show excellent agreement with the numerical simulations presented in Fig. 3b. At normal incidence, near 0.56 THz, destructive interference between the two modes results in a transmission close to 0.98. As the incident angle increases, a distinct spectral splitting can be observed, attributed to the oblique incident wave vector processing an in-plane wave vector component that couples with the continuum, causing BIC mode leakage. Isolated quasi-BICs can also be observed at frequency points of 0.47 THz and 0.67 THz, but these are beyond the scope of this work. It is worth noting that the transmission peaks and dips only reflect the Hermitian part of the Hamiltonian matrix, which is typically damped by losses and therefore does not correspond to the eigenfrequency points. The eigenfrequency points we discuss correspond to the poles of the *S*-matrix. The spectral line at the EP typically resembles a line shape of electromagnetic-induced transparency (EIT)[20]. Repeating the fitting process for transmission spectra measured at different angles yields two modes' complex eigenfrequencies. Fig. 3c shows the calculated and fitted quality factors. Due to material losses, the experimentally measured maximum Q-factor is approximately 2000 (compared to the simulated value of 3997), which is of the same order of magnitude as the highest Q-factor measurable in the THz regime[22]. At the EP, a sudden drop in the Q-factor to 106 is observed, after which it stabilizes. In the supplementary materials, we provide a comparative analysis excluding material losses, showing that even a slight loss can significantly reduce the Q-factor at the BIC. In Fig. 3d, we plot the relationship of the complex frequencies of the two modes, showing that their positions tend to converge at the EP.

After describing the transmission characteristics of BICs transitioning to EPs with varying incident angles, we further demonstrated the active switching capability of EP through optical pumping. Specifically, we maintained the incident angle of the THz wave at $\theta = 1.5°$ and used a 980-nm continuous-wave laser to optically pump the silicon sample, thereby increasing the loss of the silicon through photogenerated carriers. The schematic diagram of the experimental setup is shown in Fig. 4a. Fig 4b provides a clear comparison between the experimental and simulated transmission spectra. The carrier concentration for simulations is calculated based on the given pump power as detailed in Section 6 of the Supplementary Material. With an optical pump power changing from 0 mW (dark blue curve) to 500 mW (dark red curve), the corresponding optically doped carrier density increases from $1 \times 10^{13}/cm^3$ to $7 \times 10^{14}/cm^3$. Consequently, the strong damping caused by the photoinduced free carriers disrupts the high Q-factor sustained in the silicon resonators. As the pump power increases further, a substantial number of photocarriers dominate the decay process, leading to a broadening of the resonant modes. This indicates that the high-Q EP state of the system is disturbed. To provide a more intuitive description of the pumping modulation, we define the transmission difference, $\Delta t$, as the difference between the maximum and minimum transmission at a relatively lower frequency near 0.56 THz, i.e., $\Delta t = t_{max} - t_{min}$. Fig. 4c shows the variation of $\Delta t$ under different pump powers. The change in $\Delta t$ from 0.79 to 0.38 indicates a modulation depth of about 52%. Furthermore, through numerical simulations and data fitting, we plotted the Q-factor and the difference in the imaginary part of the eigenfrequency as functions of carrier concentration. It was found that the external optical pump primarily introduces losses as previously mentioned, which break the mode's degeneracy of the system. The carrier concentration required for the optical pump modulation of our BIC-derived EP is an order of magnitude lower than that required when only considering the quasi-BIC[21].

Another significant feature of the EP is that a 2π-phase accumulation can be achieved when encircling its parameter space, providing more options for wavefront control[36-38]. At 0.56 THz, the simulated transmission without laser pumping in Fig. 5b shows that the metasurface with the smallest radius of 64.0 μm, achieves a phase of −119.0°, while the unit with the largest radius of 71.7 μm, reaches a phase of 150.1°∘, resulting in a total phase coverage of 269.1° and with all units maintaining an amplitude above 0.65, these combined features confirm the system's capability for implementing effective 2-bit beam steering. After pumping shown in Fig. 5c, however, the phase coverage narrows to 65° with an average amplitude of just 0.32, making effective wavefront manipulation unattainable. According to the simplified generalized Snell's law: $\sin\theta_{\text{def}} = \frac{\lambda}{2\pi} d\varphi/dx$, where $\theta_{\text{def}}$ is the theoretical deflection angle, λ is the operating wavelength, and $d\varphi/dx$ is the phase gradient between adjacent structural units. Thus, we can obtain the theoretical deflection angle is about 35.6°. We performed beam deflection experiments using the setup shown in Fig. 5a. A THz beam was incident at 1.5° onto the EP phase gradient metasurface, with the transmitted THz signals collected by a receiving antenna scanning the angle behind the metasurface. As shown in Fig. 5c, a signal at 0.56 THz was detected at a deflection angle of 35.6°, demonstrating excellent agreement with theoretical predictions. However, as shown in Fig. 5d, after optical pumping with a power of 500 mW, the deflected THz signal was reduced to 33.74% of its level before pumping, indicating that the system was no longer in the EP state. Our experiments have demonstrated that the structure, designed around the EP, not only functions as a beam deflector on gradient metasurfaces with varying radii, but also highlights its potential as an optical switch leveraging the phase mechanism associated with the EP.

## Discussion

In summary, we have demonstrated the phenomenon of the transition from one singularity to a two-dimensional ring in non-Hermitian systems, and through the experimental manipulation of the incident wavevector angle, we were able to observe the transition from BIC to EP in metasurfaces, providing new insights into the interplay between these two topological singularities. Additionally, optical pumping can modify the carrier concentration in silicon, breaking the system's degeneracy and enabling dynamic switching of the EP. Through this switching mechanism, we utilized the EP phase singularity to achieve a THz transmission beam deflector switch. The BIC-derived EP presented here offers exciting possibilities for realizing compact on-chip high-sensitivity sensing, dynamic wavefront control, and other applications in non-Hermitian physical systems. Our proposal is a general approach that can be extended to optical or microwave regimes, as long as the discussion involves non-Hermitian systems with two or even multiple mode coupling. Although we chose to shift coupling by introducing oblique incidence, which could potentially be used for designing an angle detector, similar results are expected under normal incidence if the system's rotational symmetry is broken.

## Materials and methods

**Fabrication:** First, a 400 μm thick silicon wafer is thinned down to 120 μm and then diced into pieces measuring 2×2 cm$^2$. After that, the sample surface was spin-coated with AZ10XT at a speed of 2000 cpm for 45 seconds, followed by a soft bake at 110°C for 120 seconds, resulting in a thickness of approximately 9.4 μm. Subsequently, a photolithographic pattern was created to serve as an etching mask. A deep reactive ion etching (DRIE) process is performed, resulting in patterned areas of 1.8×1.8 cm$^2$, covering a large area of free-standing metasurfaces consisting of more than 6,000 units.

**Numerical Simulations:** The calculation of the band structure was performed using the eigenmode solver in the commercial software COMSOL Multiphysics, where the unit structure was modeled and Floquet periodic boundary conditions were applied in the x and y directions. The transmission was obtained by applying unit cell boundary conditions using the finite integral frequency domain solver provided by the commercial software CST Microwave Studio. The parameters for the high-resistivity silicon in the non-optically doped region used a permittivity of 11.7 and a conductivity of 0.1 S/m. The dielectric properties of the optically doped silicon were determined using the Drude model, incorporating carrier density and scattering rate of the experimental data.

**Characterization:** Measurements are performed using a fiber-based THz-TDS system, the transmission coefficient $t(\omega)$ can be obtained by $t(\omega) = |E_{sam.}/E_{ref.}|$, where $E_{sam.}$ and $E_{ref.}$ are the Fourier transforms of the time-domain electric field measurements of the sample and reference (air), respectively. A set of convex lenses is installed before and after the sample to ensure that the incident THz wave is approximately parallel. The entire assembly of the device is mounted on an electrically controlled rotating platform with a rotation range of 360° and a rotation precision of 0.1°, allowing for angle-resolved testing from normal to oblique incidence. To perform optical doping on the sample, we use a continuous wave (CW) diode laser with a wavelength of 980 nm. The beam is collimated and expanded to cover an area larger than the 1.8×1.8 cm² sample size, ensuring that the entire feature area of the sample is optically doped. The pump laser is incident at an angle of about 30 degrees onto the sample surface. The laser power incident on the sample is controlled by current, with an output laser power ranging from 0 to 500 mW.

## Data availability

The data that support the findings of this study are available from the corresponding author upon reasonable request.

## Code availability

The custom codes support the current study are available from the corresponding authors on request.

## Acknowledgments

This work was supported by the National Nature Science Foundation of China (NSFC) (Nos. 62222106, 62288101, 62027807, 92163216, 92463308, 62071217, and 62275118), and the Fundamental Research Funds for the Central Universities. The authors also thank Xinzhan Semiconductor Equipment (Shanghai) Co., Ltd. for their contributions to the thinning of silicon wafers.

## Author contributions

LW and HL performed the simulations and prepared the samples; JWL, AXL, and JLH helped on experimental spectrum measurements; HD and QNL helped on the measurement optical pump setup; JBW, HBW, BBJ, JC, and PHW provided a constructive discussion on the idea and the manuscript. LW analyzed the data. LW, HL, CHZ, and KBF wrote the manuscript; CHZ and KBF supervised the project.

## Competing interests

The authors declare no competing interests.

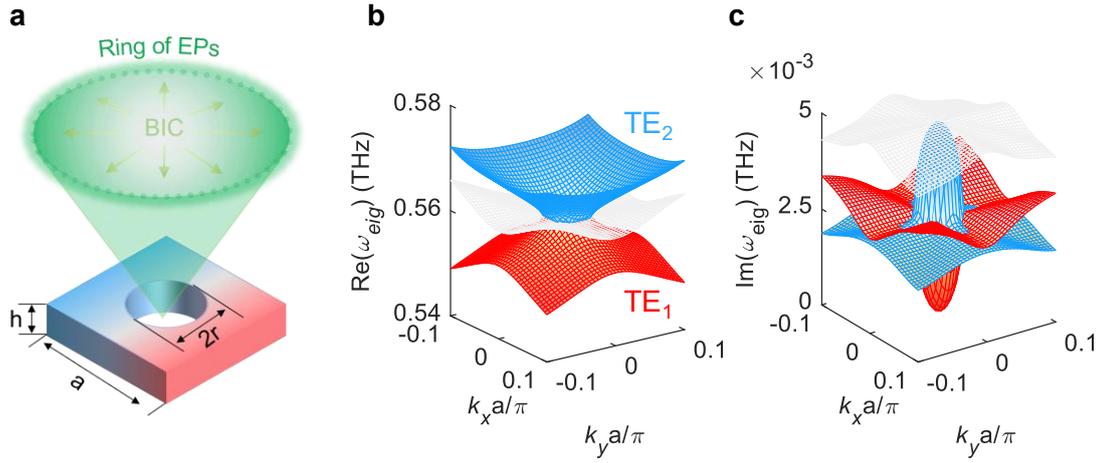

**FIG. 1. Transition from one bound singularity to a two-dimensional exceptional ring. a**, Schematic of the transition of BIC to EP ring in a designed unit cell. Structural parameters: period a = 230 μm, hole radius r = 70.4 μm, thickness h = 120 μm. **b**, The real parts of the calculated TE-like eigenmodes in *k*-spaces. **c**, The imaginary parts of the calculated TE-like eigenmodes in *k*-spaces. The red surface represents the $TE_1$ mode, while the blue surface corresponds to the $TE_2$ mode.

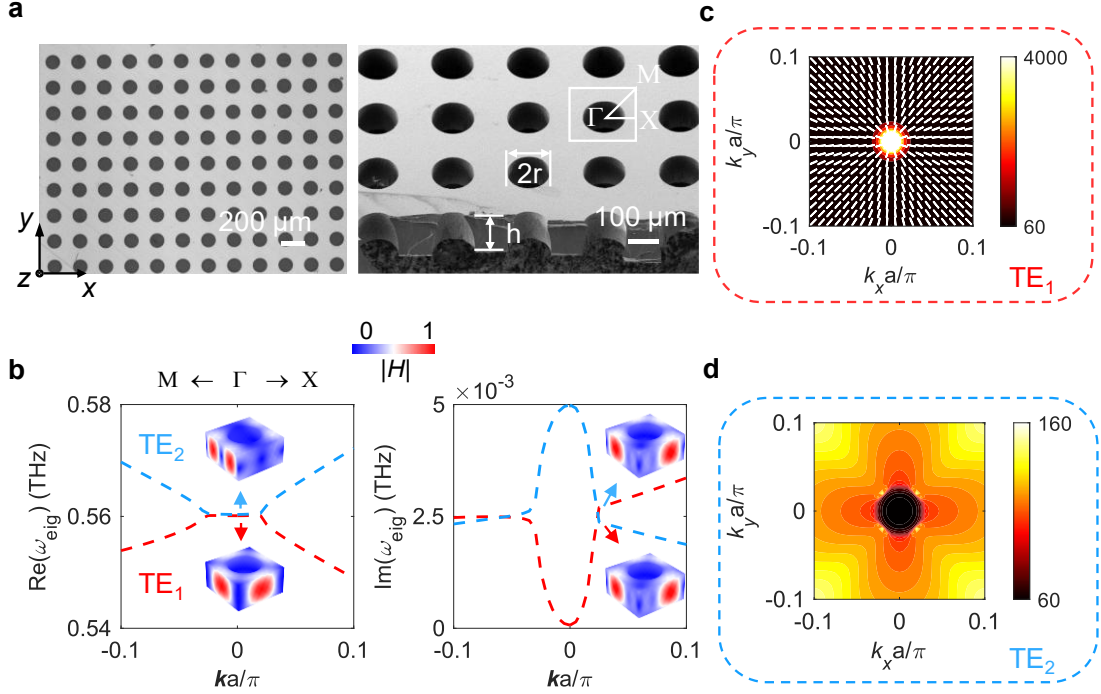

**FIG. 2. EP and FW-BIC were observed in the calculated TE-like band structure. a**, Left: Optical microscope image of the prepared sample supporting the transition from FW-BIC to EP. Right: SEM image of the cross-section of the sample. **b**, Detailed views of the real and imaginary parts of the two discussed eigenmodes are provided. The insets depict the normal magnetic field distributions of the two modes at the Γ point (BIC) and away off-Γ point (EP). **c, d**, The calculated Q-factors of the $TE_1$ and $TE_2$ mode, with the white ellipse indicating the far-field polarization around the BIC.

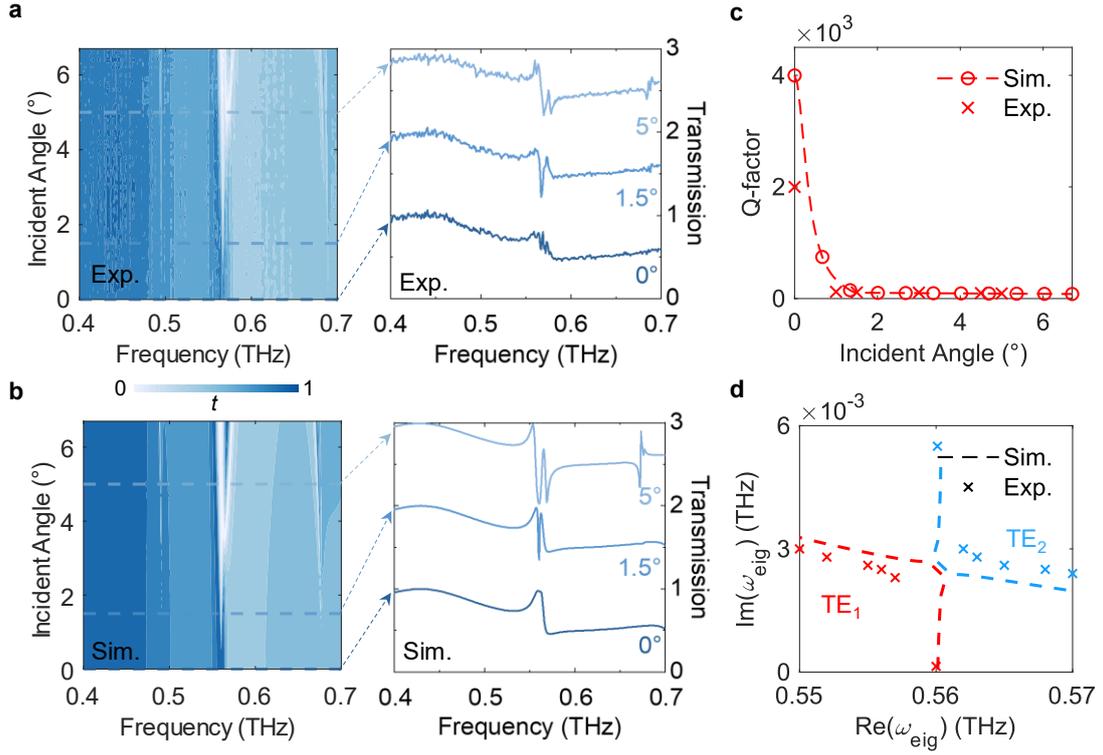

**FIG.3. Experimental demonstration of the transition process from FW-BIC to EP. a**, **b**, Left: the color maps show the measured (**a**) and simulated (**b**) transmission spectra as a function of the incident terahertz angle. Right: the extracted transmission spectra at angles of 0 °, 1.5 °, and 5 ° are detailed, corresponding to FW-BIC, EP, and a more quasi-BIC, respectively. **c**, The variation of the calculated Q-factor with the incident angle. **d**, The real and imaginary parts of the eigenfrequency exhibit a trend of approaching each other near the EP.

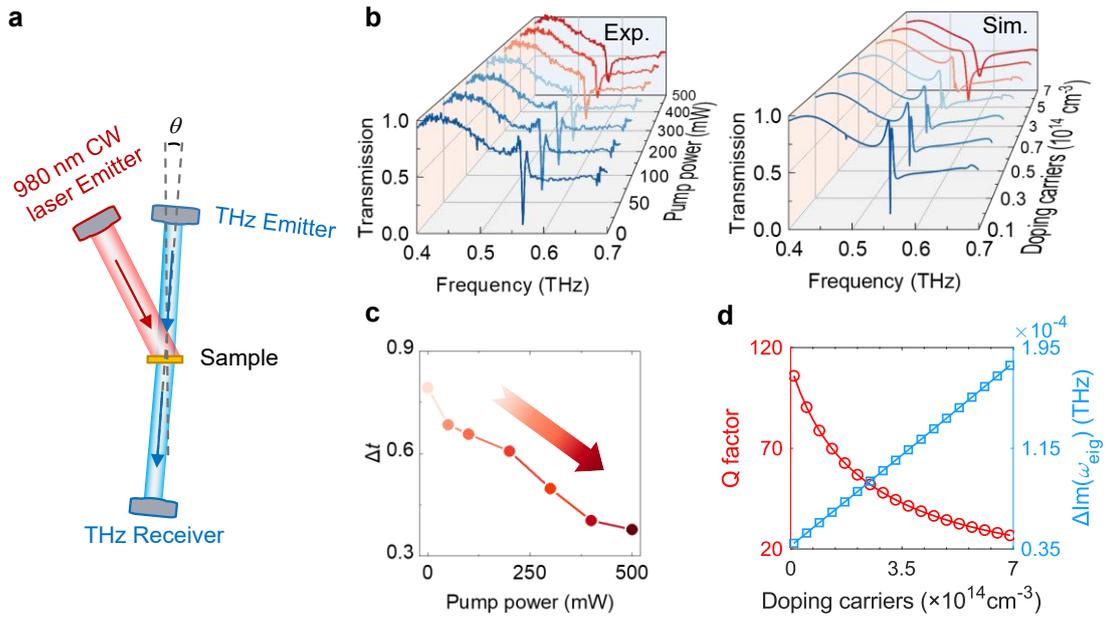

**FIG 4. Results of the optical pumping experiment at the EP point. a**, Schematic diagram of the experimental setup. **b**, Left: experimental data demonstrating the pumping process through changes in pump power, right: simulated data demonstrating the pumping process through changes in carrier concentration. **c,** Dependence between transmission difference $\Delta t$ and pump power. **d**, Variation of the Q-factor and the degree of nondegeneration of the imaginary part of the two modes with carrier concentration.

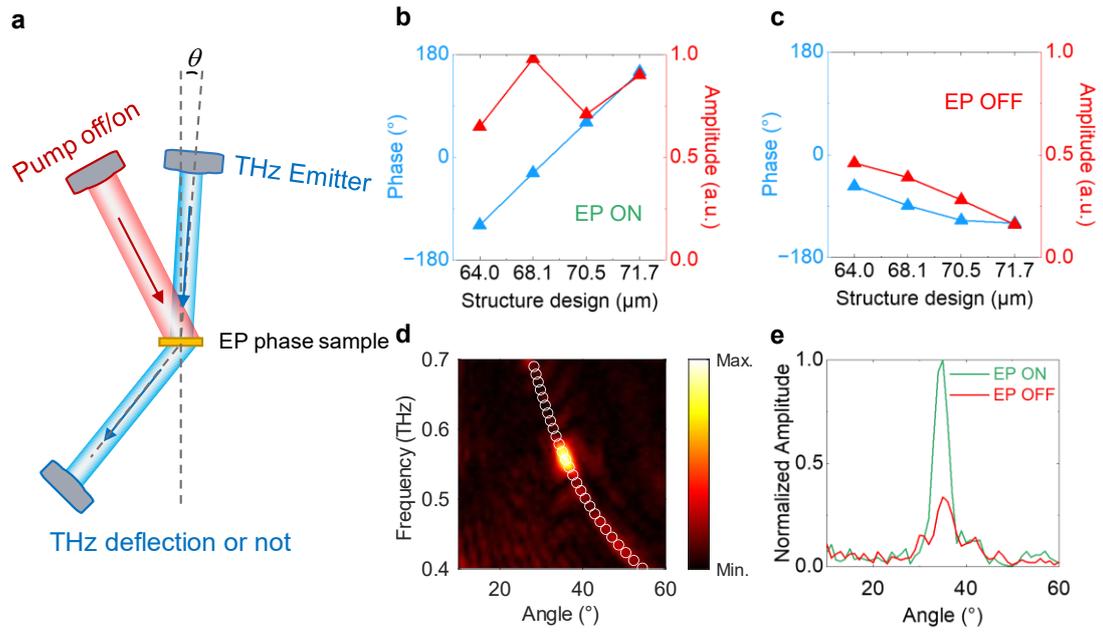

**FIG.5 Switch based on EP as a phase singularity. a**, Schematic diagram of the experimental setup for THz beam deflection. **b**, **c**, Amplitude and phase diagrams for the structural units with four radius parameters before and after the pump (carrier concentration is $7 \times 10^{14}$ cm$^{-3}$, corresponding to the pump power of 500 mW) at the EP point. **d**, Experimental measurements of the beam deflection result, with white circles indicating theoretical predictions. **e**, Normalized THz signal amplitude results at different angles; green line represents before the pump (EP on), and red line represents after the pump (EP off).